\PassOptionsToPackage{dvipsnames,table}{xcolor}
\documentclass[9pt,shortpaper,twoside,web]{ieeecolor}
\usepackage{generic}

\usepackage{cite}
\usepackage{hyperref}
\usepackage{textcomp}
\usepackage{graphicx}
\usepackage[utf8]{inputenc}
\usepackage{xfrac}
\usepackage{subcaption}
\usepackage[switch]{lineno}
\usepackage{amsmath,amssymb,amsfonts}
\usepackage{algorithmic}
\usepackage{url}
\usepackage{multirow}
\usepackage{float}
\usepackage{booktabs}
\usepackage{xcolor}

\def\BibTeX{{\rm B\kern-.05em{\sc i\kern-.025em b}\kern-.08em
    T\kern-.1667em\lower.7ex\hbox{E}\kern-.125emX}}
\newcommand{\yield}{N_\gamma}

\newcommand{\Vov}{V_{ov}} 

\begin{document}

\title{Sources of Internal Crosstalk in Silicon Photomultipliers}
\author{L. Wang\thanks{All authors are with TRIUMF, Vancouver, British Columbia V6T 2A3, Canada (e-mail: lwang2@triumf.ca).}, 
        H. Lewis, 
        F.~Reti\`ere, A. de St Croix\thanks{A. de St Croix is also with Queen's University, Kingston, ON K7L 3N6, Canada}}

\maketitle
\begin{abstract}
Silicon Photomultipliers (SiPMs) have been widely adopted for photon detection in next-generation dark matter and neutrino detection experiments. Internal crosstalk, resulting from secondary photons produced during avalanche multiplication, \textit{is a significant noise mechanism in SiPMs.}
This work presents experimental data for trends in internal crosstalk probability with temperature and overvoltage for two different SiPM devices, and demonstrates a novel method for determining the region within the SiPM in which the primary carriers for crosstalk avalanches originate. This is done by using measurements of avalanche triggering probabilities for different charge carrier types to identify the device regions in which crosstalk avalanches are produced. This is possible because secondary photons absorbed in n-doped regions will produce hole-triggered avalanches, and those absorbed in p-doped regions will produce electron-triggered avalanches. The relative probabilities of different optical transmission paths for secondary photons can therefore be evaluated, and  the effectiveness of existing crosstalk mitigation methods for each optical pathway can be assessed. We identify optical reflections from the device surface as a significant source of crosstalk, likely to dominate in devices with effective crosstalk mitigation in bulk.
\end{abstract}

\begin{IEEEkeywords}
Silicon (Si), Silicon Photo\-multiplier (SiPM), Correlated Signals,
Internal Crosstalk, Direct Crosstalk, Delay\-ed Crosstalk
\end{IEEEkeywords}


%
%
%
%
\section{Introduction}
\label{sec:intro}

Silicon photomultipliers are used as photon counters in many particle physics applications \cite{gallina_performance_2022, carnesecchi_light_2020}, as well as in medical imaging \cite{lecoq_sipm_2021} and ranging applications \cite{4907333, 7049394, Bruschini2019}. Their advantages include mechanical robustness, fast timing, low operating voltage and insensitivity to magnetic fields. SiPMs consist of an array of single photon avalanche diodes (SPADs), which are operated at a bias voltage above their characteristic breakdown voltage such that the absorption of a single photon will generate a measurable pulse of current, referred to as an `avalanche'. Pixelation allows multiple photons to be detected simultaneously with single photon resolution. Analog SiPM structures sum the outputs of each pixel. The amplitude of the overall output signal is ideally proportional to the number of incident photons, but may also be affected by various correlated noise mechanisms.

One significant noise mechanism in SiPMs is optical crosstalk \cite{hampel_optical_2020}. During an avalanche, relaxation of charge carriers accelerated by the high electric field in the device results in the emission of secondary photons \cite{akil_multimechanism_1999,mclaughlin_characterisation_2021} Absorption of these photons by other SPADs within the same SiPM (`internal' crosstalk, iCT), or by other SiPMs within a larger detector system (`external' crosstalk, eCT), results in noise avalanches which impact the photon counting resolution of the detector. Photons which escape from the SiPM and are reflected back to other microcells within the same device are also classified as eCT. Crosstalk may also be divided into prompt or `direct' (DiCT) and `delayed' (DeCT) \cite{NAGY201444, 7102791}; DiCT is generally considered to be more problematic as it cannot be distinguished from genuine, prompt multi-photon events, whereas DeCT events may be identified by their pulse shape \cite{tajima_studies_2023}.

The objective of this study is to analyze the source mechanisms of internal crosstalk, with the aim of understanding how crosstalk can be mitigated in future devices. Using experimental data we relate the crosstalk characteristics of different devices to details of the modelled device structures determined in \cite{croix2025mappingphotondetectionefficiency}. ‘Frontside-illuminated SiPMs have been reported with various crosstalk mitigation methods \cite{yamamoto_recent_2019,oleynikov_after-pulsing_2017}, as well as devices which combine mitigation technologies to produce very low crosstalk rates \cite{merzi_nuv-hd_2023}, but future developments of advanced device structures such as backside-illuminated (BSI) architecture will require considerable engineering effort to reduce crosstalk to an acceptable level. BSI SiPMs are desirable for their improved PDE and suitability of 3-D digital integration \cite{ninkovic_avalanche_2007,parellada_monreal_back_2021}, but current structures are limited by high crosstalk rates \cite{parellada_monreal_back_2021,van_sieleghem_backside-illuminated_2022}. We present a tool which can be generalized to different SiPM structures, feeding into an iterative design process and accelerating the development of SiPMs optimized in all performance metrics. Incorporating all available mitigation techniques into a novel device structure increases device complexity and relevant technologies may not be available to all manufacturers. The presented methodology can be used to discriminate between sources of crosstalk and determine what mitigation methods are necessary and appropriate for a given device structure.

The methodology is to determine the region within the device where different crosstalk avalanches originate, illustrated using two representative p-on-n SiPM architectures. This is done by comparing trends in crosstalk probability against overvoltage with the known trend in the respective avalanche triggering probabilities for events initiated by electrons or holes, allowing the dominant initiating carrier type to be identified. In addition, DeCT is used to make a direct experimental measurement of the hole-initiated avalanche triggering probability, as electrons contribute negligibly for delayed events. This is the first such measurement in a p-on-n SiPM and will be useful in breaking degeneracies in characterization \cite{croix2025mappingphotondetectionefficiency}.

Crosstalk probability has been measured as a function of overvoltage and temperature (from -18C to -110C) for two p-on-n SiPMs: the FBK VUV-HD3 \cite{CAPASSO2020164478} and the Hamamatsu VUV4 \cite{hamamatsu_photonics_vuv-mppc_2017}. These devices differ in structure, with photodetection efficiency modelling indicating a deeper P-N junction for the Hamamatsu device \cite{croix2025mappingphotondetectionefficiency}. The Hamamatsu device also contains metal-filled trenches intended to mitigate crosstalk \cite{oleynikov_after-pulsing_2017, hamamatsu_photonics_vuv-mppc_2017}.
Section \ref{sec:methodology} describes the experimental and analysis methodology and our model for iCT rate, \autoref{sec:iCT} contains an assessment of the crosstalk characteristics of the two devices and a description of crosstalk photon and carrier propagation, and \autoref{sec:source_analysis} contains an analysis of the crosstalk sources in each device. Determining the variation, if any, in crosstalk probability with temperature is a secondary motivation for this work - this is intended to support mass testing of these devices at the production stage of a liquid Xenon (LXe) experiment, by determining whether measurements taken at higher temperatures (-20C to -40C) are sufficient to characterize the crosstalk performance of devices at the LXe temperature of -110C. The two devices tested are the candidates for inclusion in the nEXO experiment, which will search for neutrinoless double beta decay in LXe [REF]. Previous studies have tested the temperature dependence of crosstalk probability in similar SiPMs \cite{acerbi_cryogenic_2017}, but this has not yet been characterized for these devices. 

\section{Experimental methodology and iCT model}
\label{sec:methodology}
Measurements were taken in vacuum using the VERA setup, which is described in detail in \cite{lewis_measurements_2025}. Device output pulses were converted to a voltage reading using a high-speed transimpedance amplifier constructed in-house, and waveforms from the amplifier were recorded using a CAEN DT5730 digitizer. Cooling was provided by liquid nitrogen supplied to a cold finger, with feedback control used to maintain temperature stability of ±0.1 K. All measurements in this study were taken in dark. 

The mean number of iCT avalanches produced per primary pulse, $\lambda_{iCT}$, can be described using an effective model as a function of overvoltage $\Vov$ as given in \autoref{eq:lambdaiCT}.


\begin{equation}
    \lambda_{iCT}(\Vov) = \yield(\Vov)\big[ P_{A,p}P_e(\Vov) + P_{A,n}P_h(\Vov) \big]
    \label{eq:lambdaiCT}
\end{equation}
The term in square brackets represent an `internal detection efficiency' for the secondary photons generated by the primary SPAD avalanche. This is similar to the approach validated in \cite{croix2025mappingphotondetectionefficiency} where:

\begin{itemize}
    \renewcommand{\labelitemi}{$\cdot$}
    \item $\yield(\Vov)$ - the mean number of secondary photons emitted per avalanche. This parameter is reported to increase linearly with the charge gain of the device \cite{Raymond_stimulated_emission_TED_2024,lewis_measurements_2025}, with an emission spectrum peaked in the near-IR. As such $\yield(\Vov)$ can be represented by a linear function, which simplifies to:

\begin{equation}
    \yield = k\cdot Gain= \frac{kC}{q_{e}}\Vov 
    \label{eq:kV}
\end{equation}
    Where $C$ is the capacitance of a single SPAD microcell and $q_{e}$ is the electron charge. The term $\frac{C}{q_{e}}\Vov$ defines the gain of SiPM at a certain over voltage. Then $k$ represents the number of secondary photons emitted per carrier during avalanche. 

\item $P_{A, p}$ and $P_{A, n}$ - an effective parameter which represents the combination of the average photon absorption probability in the p (n) region of the device and carriers' collection efficiency, which is the probability that the photo-generated electron (hole) reaches the high field region from the site where it is generated. Microscopically $P_A$ depends on each generated photon's wavelength, production location and propagation path through the device. These variables are averaged over in $P_A$, which is the convolution of the emitted photon's wavelength distribution, direction, path, and absorption probability throughout the 3D structure. It should be noted that $P_a$ is not calculated as part of the analysis, but instead composes part of the $\alpha$ and $\beta$ values, described below, which will be used to determine the relative proportions of crosstalk events that are electron or hole generated.
    \item $P_e(\Vov)$ and $P_h(\Vov)$ - the respective probabilities that an electron or hole reaching the high-field region of the device triggers an avalanche. 
\end{itemize}

In the equation, only the first term follows a Poisson distribution, while the latter two terms adhere to binomial distributions. Assuming the junction volume and therefore $P_A$ is independent of $\Vov$, ~\autoref{eq:lambdaiCT} is divided by gain such $P_{e,h}(\Vov)$ are the only terms with $\Vov$ dependence, defining the reduced internal cross-talk $\lambda_{iCT}^*$ as below:

\begin{align}
    \lambda_{iCT}^* = \frac{\lambda_{iCT}(\Vov)}{Gain} &= k\big[ P_{A,p} \cdot P_e(\Vov) + P_{A,n} \cdot P_h(\Vov) \big] \\
     & = \alpha P_{e}(\Vov) + \beta P_{h}(\Vov)
    \label{eq:lambdatoV}
\end{align}
Where $\alpha$ is defined appropriately as $\alpha \equiv k \cdot P_{A,p}$, proportional to the average photon absorption in the effective p-region. Similarly for $\beta$ and absorption in the effective n-type region. This equation transforms the internal cross-talk to a linear combination of $P_{e}(\Vov)$ and $P_{h}(\Vov)$ which can be fit for $\alpha$ and $\beta$, estimating the relative contribution to cross-talk from carriers generated in either the p or n region.

\section{Internal crosstalk performance of FBK and HPK SiPMs}\label{sec:iCT}

\begin{figure}[htbp]
\centering
\begin{subfigure}[b]{0.4\textwidth}
    \centering
    \includegraphics[width=\textwidth]{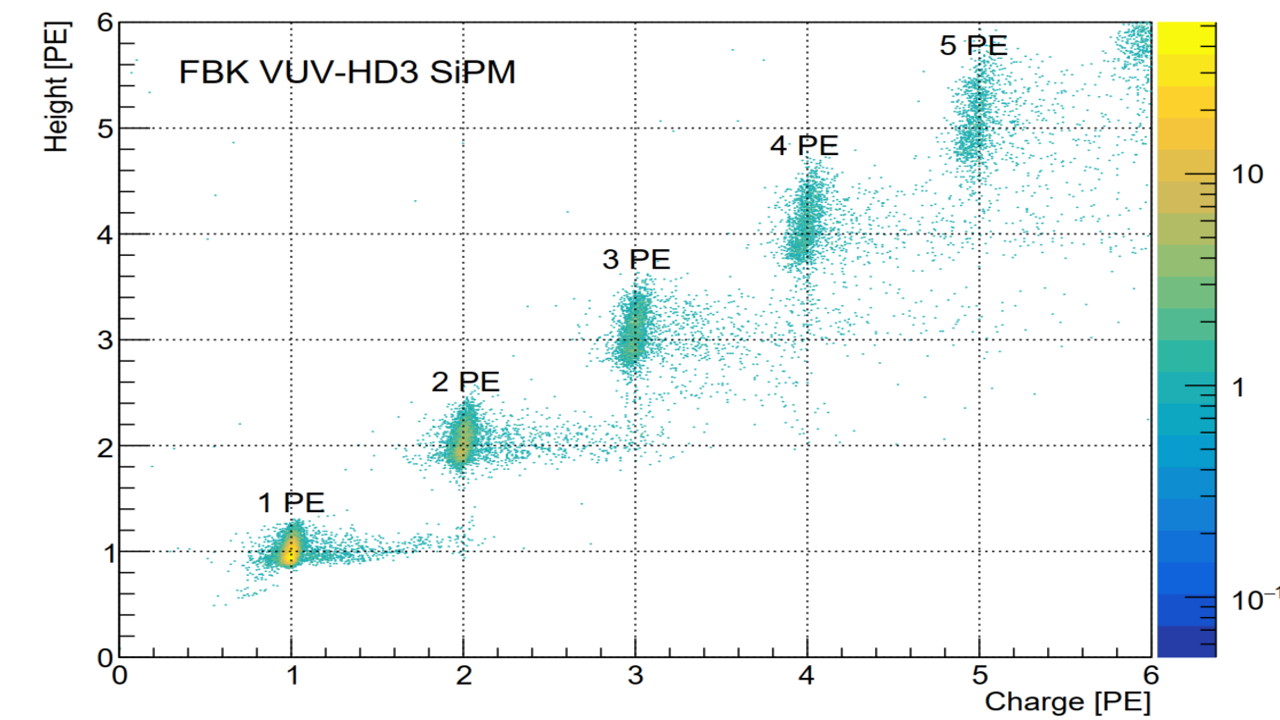}
    \caption{2D distribution of pulse height versus charge for FBK VUV-HD3.}
    \label{fig:fbk_height}
\end{subfigure}
\hfill
\begin{subfigure}[b]{0.4\textwidth}
    \centering
    \includegraphics[width=\textwidth]{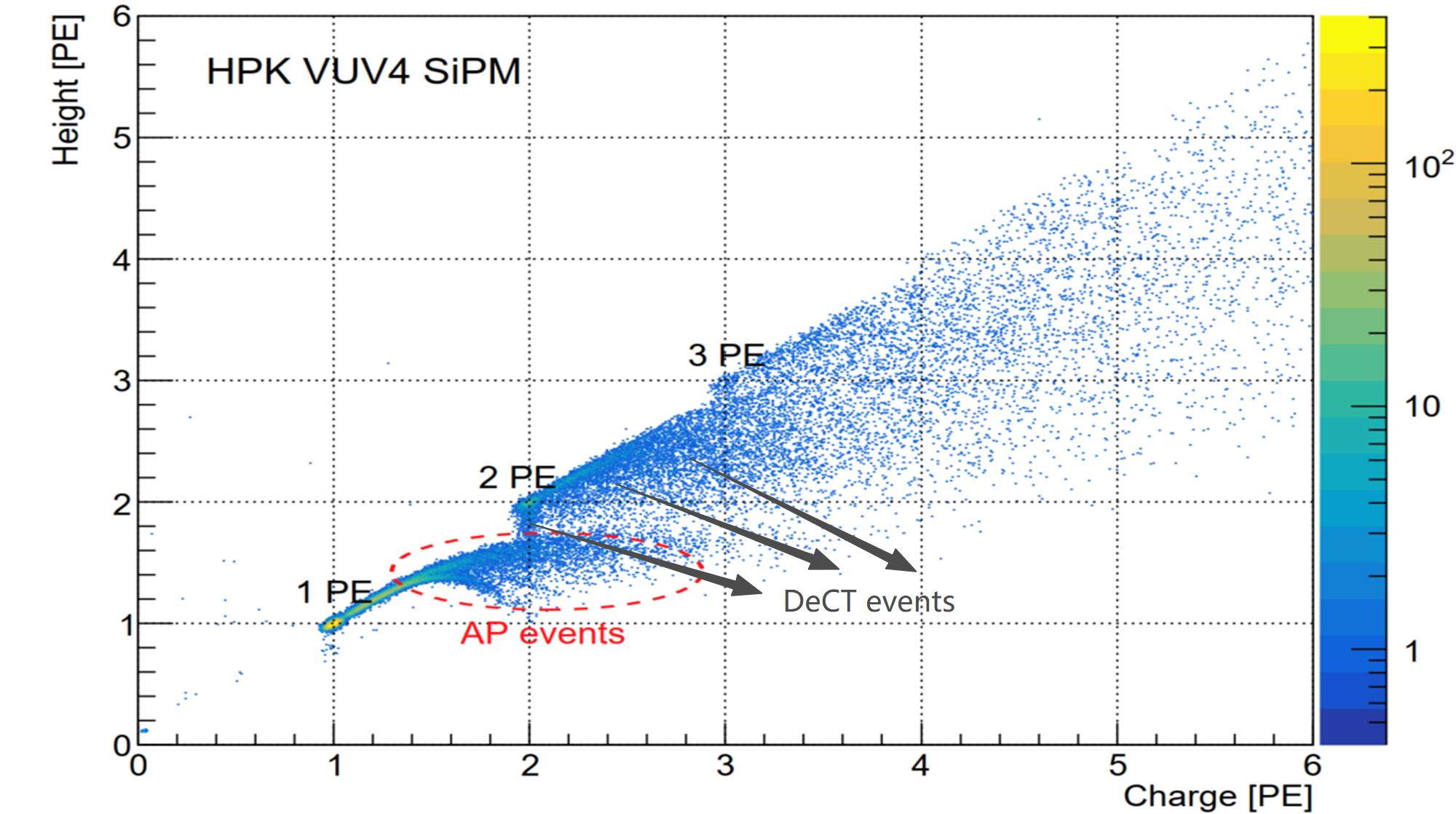}
    \caption{2D distribution of pulse height versus charge for HPK VUV4.}
    \label{fig:hpk_height}
\end{subfigure}

\vskip\baselineskip

\begin{subfigure}[b]{0.4\textwidth}
    \centering
    \includegraphics[width=\textwidth]{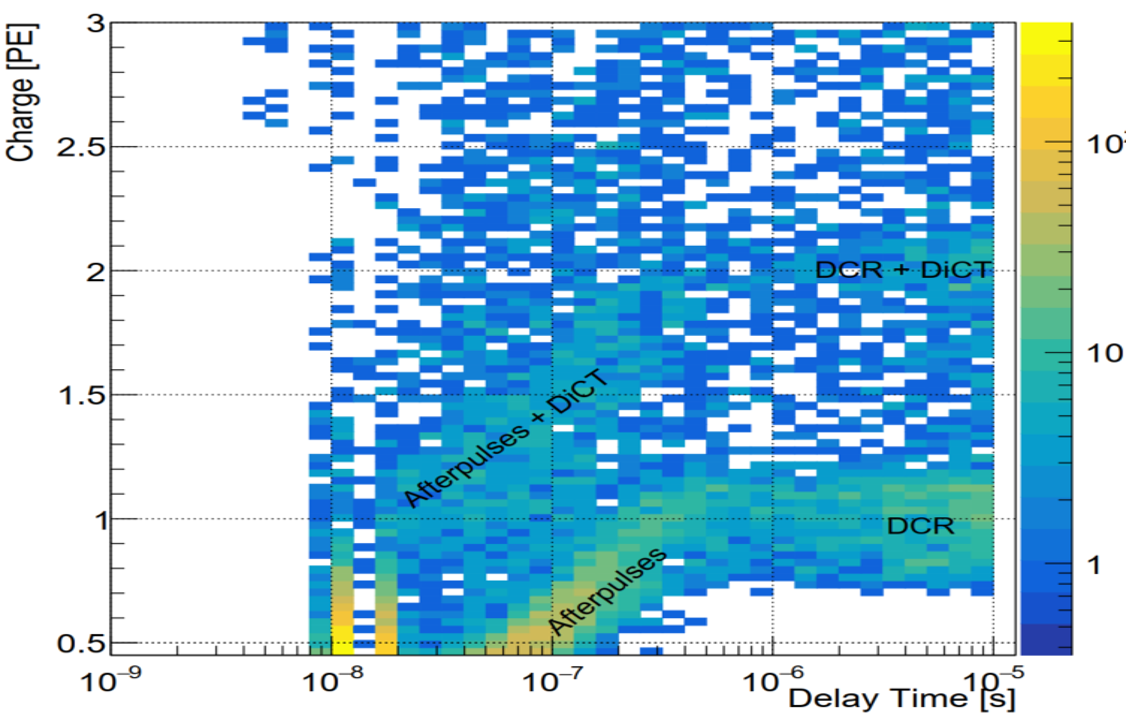}
    \caption{2D distribution of charge versus delay time for the FBK VUV-HD3}
    \label{fig:fbk_time}
\end{subfigure}
\hfill
\begin{subfigure}[b]{0.4\textwidth}
    \centering
    \includegraphics[width=\textwidth]{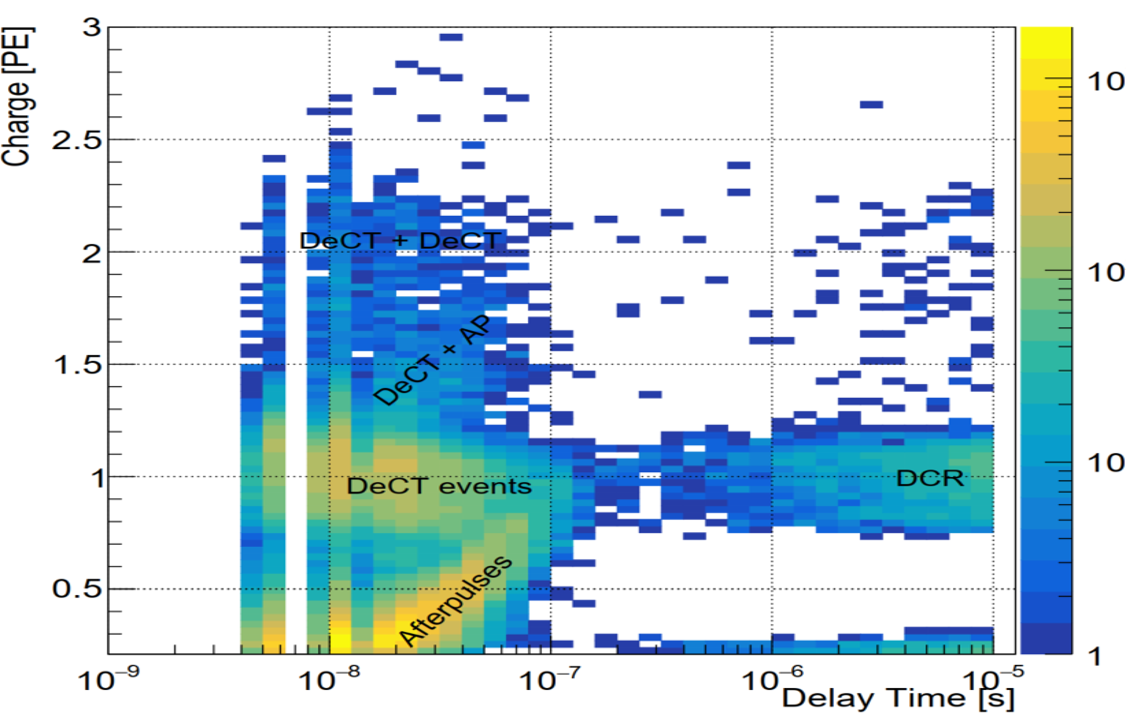}
    \caption{2D distribution of charge versus delay time for the HPK VUV4}
    \label{fig:hpk_time}
\end{subfigure}

\caption{Charge-Time and Charge-Height distributions for the Hamamatsu VUV4 and the FBK VUV-HD3.}
\label{fig:sipm_2dplots}
\end{figure}

We initially compare the iCT performance of the two devices. ~\autoref{fig:hpk_height} and ~\autoref{fig:fbk_height} show 2D histograms of charge against pulse height, and ~\autoref{fig:fbk_time} and ~\autoref{fig:hpk_time} show distributions of the secondary pulse's charge against the time elapsed since the previous pulse. These figures are commonly used for illustrating SiPM performance parameters \cite{7102791, gallina_characterization_2019_HPK}. Comparing the first two figures it can be observed that the FBK SiPM has a significantly higher number of iCT avalanches than the HPK SiPM, due to the amount of 4 PE and even 5 PE events given the same measurement conditions. However, ~\autoref{fig:hpk_height} reveals a greater number of afterpulsing and DeCT events in the HPK device, as marked by the arrows and circle. DeCT events are indicated by the area where the charge is equal to 1 PE with a measurable delay between pulses, with  ~\autoref{fig:hpk_time} showing that DeCT greatly exceeds DiCT in the HPK device. A comparison between  ~\autoref{fig:fbk_time} and  ~\autoref{fig:hpk_time} reveals that the FBK SiPM’s CT composition includes almost no DeCT events. These data indicate that the crosstalk characteristics of the two devices differ significantly, providing a useful testbed for a study of the underlying mechanisms.

\begin{figure}[htbp]
\centering

\begin{subfigure}[b]{0.5\textwidth}
\centering
\includegraphics[width=\textwidth]{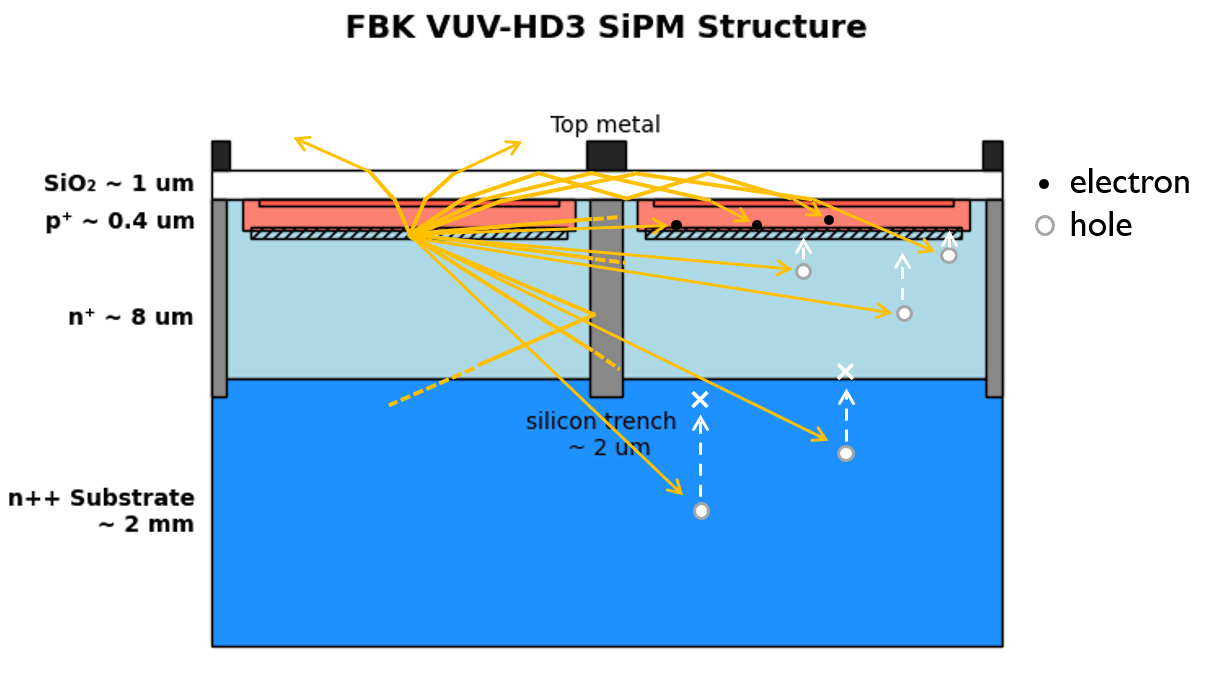}
\caption{}
\label{fig:FBK_cartoon}
\end{subfigure}

\vspace{1em}

\begin{subfigure}[b]{0.5\textwidth}
\centering
\includegraphics[width=\textwidth]{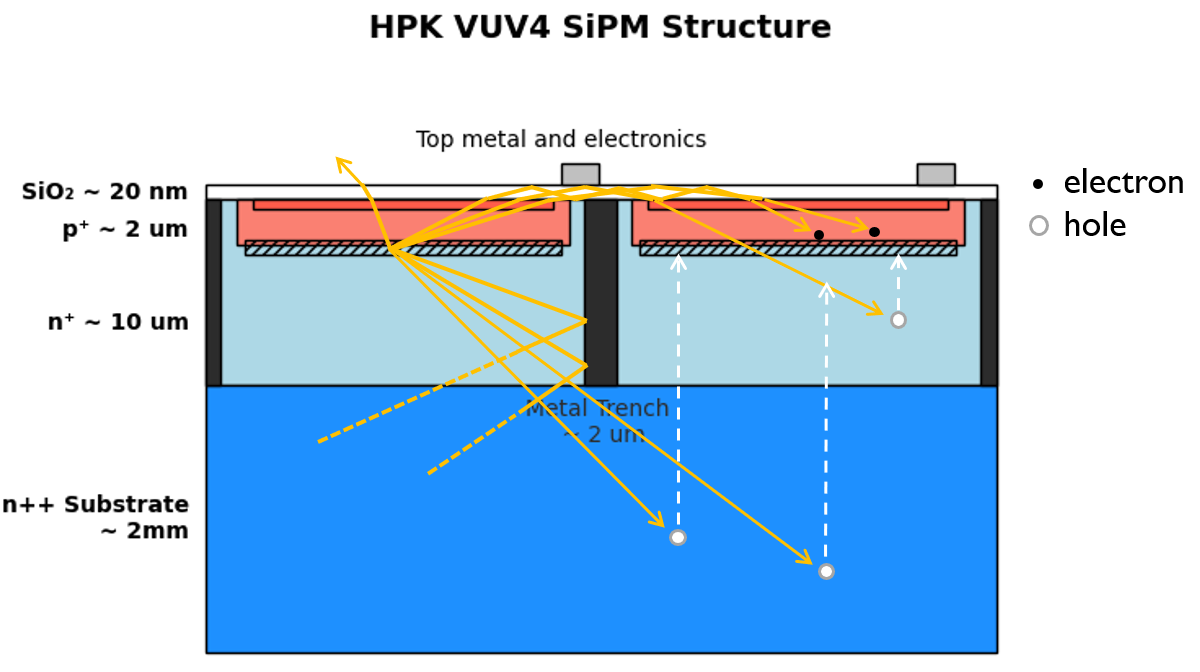}
\caption{}
\label{fig:HPK_cartoon}
\end{subfigure}

\caption{Schematic diagrams of the FBK VUV-HD3 and Hamamatsu VUV4, showing the propagation paths for secondary photons described in   \autoref{sec:iCT}. The dimensions shown here are those determined by modelling of the wavelength and angular dependence of photon detection efficiency, as described in \cite{croix2025mappingphotondetectionefficiency}. Device dimensions are not to scale, and the rough thicknesses of each layer is shown in the left-side.}
\label{fig:SiPM_cartoon}
\end{figure}

Sources of iCT depend on the absorption site of secondary photons: absorption in the p-region leads to electron-triggered DiCT; absorption in the n-region to hole-triggered DiCT. Deeper absorption in the substrate produces carrier pairs from which holes can diffuse back into the active region, constituting the primary mechanism for delayed crosstalk (DeCT). The relative intensity of these iCT populations depend on device structure and construction, which influences the optical paths of the secondary photons.  ~\autoref{fig:SiPM_cartoon} shows diagrams of the two SiPM structures and some optical paths which are discussed below. The device dimensions given in this figure, and used in the proceeding analysis, were determined by modelling of the wavelength and angular dependence of the photon detection efficiency of these devices. This method is detailed in \cite{croix2025mappingphotondetectionefficiency}, and the values are listed there.

The first optical path is direct transmission through the optical trenches which separate pixels. Given that we consider the majority of the active volume of these devices to be n-doped rather than p-doped \cite{croix2025mappingphotondetectionefficiency}, it is likely that DiCT avalanches produced by this mechanism will be predominantly hole-initiated. This hypothesis is supported by analysis of crosstalk sources in   \autoref{sec:source_analysis}.The trenches of the VUV-HD3 are filled with SiO$_2$ \cite{gola_nuv-sensitive_2019}, which is not completely absorbing and it is expected that a significant number of photons are directly transmitted to neighbouring SPADs. Conversely, the HPK VUV4 incorporates a metal filling in the optical trench \cite{gallina_characterization_2019_HPK}, greatly reducing the quantity of secondary photons passing directly from one depletion region to another. This explains the major reduction in DiCT when comparing the HPK and FBK results.

Secondly, photons may propagate directly to the bulk silicon contributing to DeCT via diffusion. Optical reflections from the back of the device and subsequent absorption in the substrate have also been proposed \cite{Rech:08} as a contributor to DeCT. However the FBK device shows a lack of the diffusion-driven DeCT in  ~\autoref{fig:fbk_time}. This is ascribed to the low-AP technology used by FBK in the VUV-HD3 ~\cite{CAPASSO2020164478, 7102791} which significantly shortens the lifetime of holes in the substrate, reducing the effective drift time to nanosecond scale \cite{7322059}. For the HPK device, the longer minority carrier lifetime in the substrate results in the higher number of DeCT and optically-induced afterpulsing events seen in  ~\autoref{fig:hpk_height} and  ~\autoref{fig:hpk_time}.

The third optical path to be considered is incidence on the SiPM surface, which consists of an SiO$_{2}$ passivation layer atop the silicon. Due to the large change in refractive index from silicon to SiO$_{2}$ and vacuum, the majority of photons are internally reflected \cite{lolx_Gallacher2025}. As the iCT's emission spectrum is in the NIR \cite{Raymond_stimulated_emission_TED_2024}, these internally reflected photons are predicted to be absorbed throughout the device, expected to contribute more strongly to DiCT than the other mechanisms. We note that crosstalk events generated by this path have been referred to in the literature as `external' crosstalk \cite{hampel_optical_2020}, but in the context of particle physics detectors, external crosstalk specifically refers to events where a secondary photon leaves a SiPM device and triggers a separate SiPM within the larger detector system.

\section{Analysis of crosstalk sources for FBK and HPK SiPMs}\label{sec:source_analysis}

The internal crosstalk probability $P_{iCT}(\Vov)$ is defined as the probability that a primary avalanche is coincident with one or more additional avalanches, experimentally quantified as the fraction of events with an observed charge corresponding to at least two photoelectrons, and is related to the average number of iCT events $\lambda_{iCT}(\Vov)$ by:

\begin{equation}
    \lambda_{iCT} = - ln(1 - P_{iCT})
    \label{eq:PiCT}
\end{equation}

\begin{figure}[htbp]
\centering
\includegraphics[width=.47\textwidth]{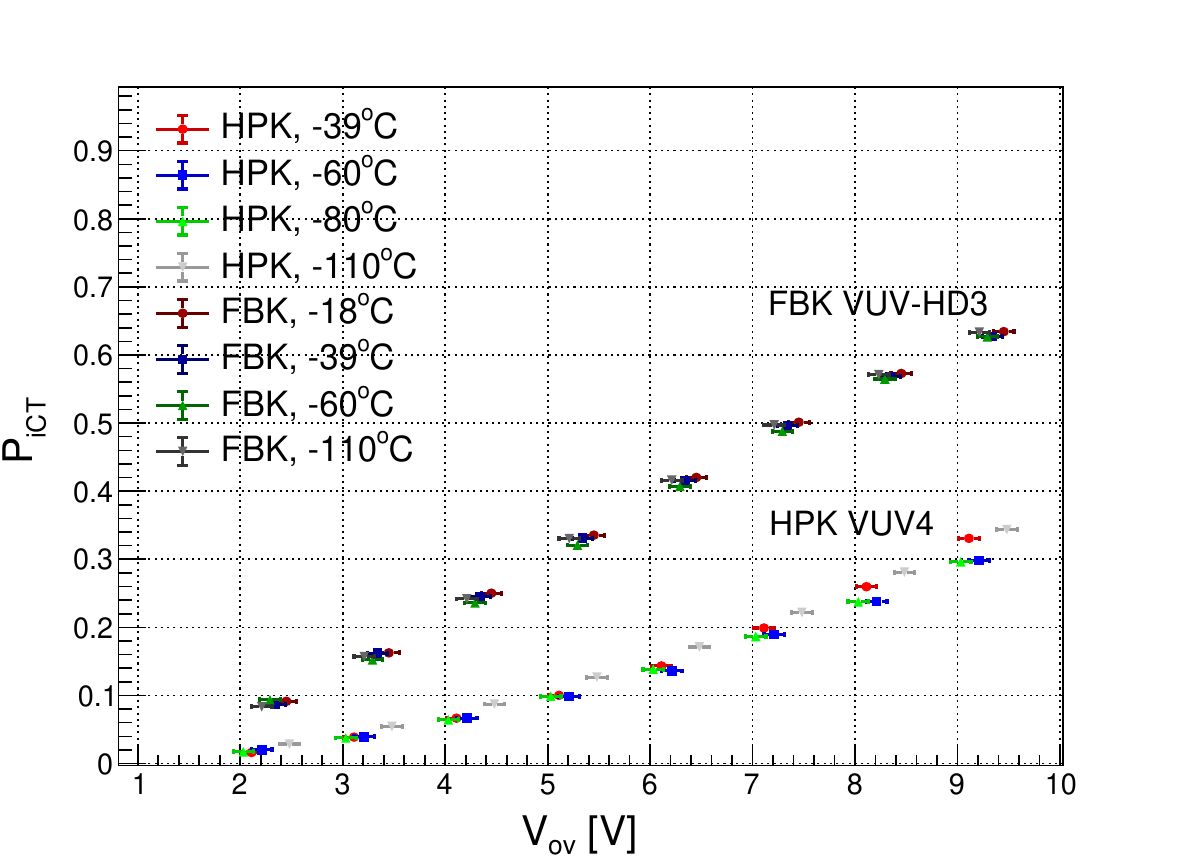}
\caption{Functions of internal crosstalk probability versus $\Vov$ for the Hamamatsu VUV4 and the FBK VUV-HD3. Four temperatures were applied in the measurement.\label{fig:CT}}
\end{figure}
\begin{figure}[htbp]
\centering
\includegraphics[width=.47\textwidth]{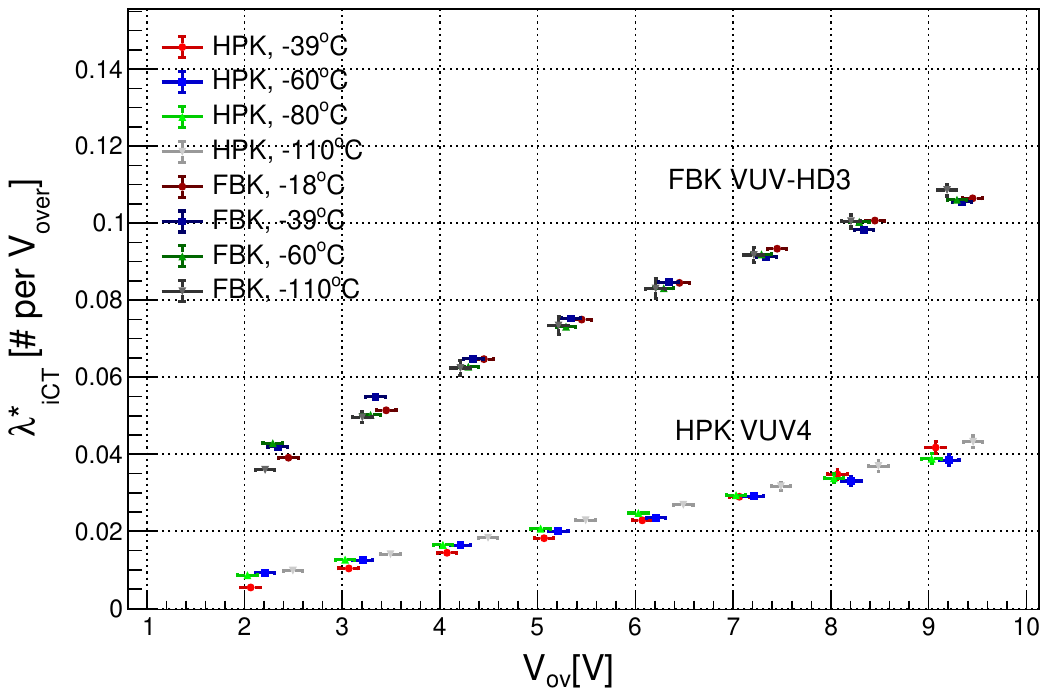}
\caption{Values of $\lambda^*$ versus $\Vov$ for the Hamamatsu VUV4 and the FBK VUV-HD3.\label{fig:CT_to_Gain}}
\end{figure}

Internal crosstalk avalanches are identified as pulses with the area greater than a single photoelectron (SPE) pulse, which is generated by the firing of a single SPAD pixel. Under dark conditions and at low temperature there is a negligible probability of multiple SPADs simultaneously avalanching due to dark noise, therefore we consider all pulses greater than $1$ PE to be generated by crosstalk. The internal crosstalk probability is therefore calculated using equation \autoref{eq:xtprob}:
\begin{equation}
    P_{iCT}(\Vov,T) = \frac{N_{1.5PE}(\Vov,T)}{N_{0.5PE}(\Vov,T)}
    \label{eq:xtprob}
\end{equation}
Where $N_{0.5PE}$ is the number of pulses with area greater than 0.5 PE and $N_{1.5PE}$ is the number of pulses with area greater than 1.5PE. 0.5PE and 1PE were selected as the threshold values, rather than 1 and 2PE, to account for any deviations in the pulse area due to noise.  ~\autoref{fig:CT} shows $P_{iCT}$ versus $\Vov$ for the two devices. Measurements were taken at 4 temperatures from -39 to -110C, with no significant temperature dependence observed. Previous studies on the temperature dependence of crosstalk in different FBK SiPM technologies have shown weak to moderate dependence of DiCT probability with respect to temperature, but this varies significantly depending on the exact technology used \cite{acerbi_cryogenic_2017,CAPASSO2020164478}. The reduced number of crosstalk events $\lambda_{iCT}^*$ ($\lambda _{iCT}$, normalized by charge gain), is shown as a function of $\Vov$ in   \autoref{fig:CT_to_Gain}.

 ~\autoref{eq:lambdatoV} can be used to relate $\lambda ^*_{iCT}$ to the proportion of avalanches initiated by electrons and holes. The trends in the electron and hole avalanche triggering probabilities with respect to $\Vov$ are known to differ significantly \cite{gallina_characterization_2019}. The electron triggering probability $P_{e}(\Vov)$ typically saturates to unity at high $\Vov$. In contrast, the hole triggering probability $P_{h}$ increases more slowly with voltage, and typically cannot be saturated at reasonable operating voltages. We describe the functional form of \( P_e \) and \( P_h \) using empirical formulae as shown in  ~\autoref{eq:Peh}:

\begin{equation}
P_{e,h}(\Vov) = 1-e^{\frac{\Vov}{V_{e,h}}}
\label{eq:Peh}
\end{equation}

This parameterization follows the method used in \cite{Ghioni1996},\cite{m_biroth_p_achenbach_w_lauth_a_thomas_analytical_2018, gallina_characterization_2019}. $P_{e,h}(\Vov)$ here is the combined expression of $P_{e}(\Vov)$ and $P_{h}(\Vov)$. $P_e(\Vov) $ has been measured independently of iCT for these devices. This is done by measuring the rate of photo-generated pulses under illumination from light where the wavelength is sufficiently short that all photons are absorbed in the p-type region of the device. Data are taken from \cite{croix2025mappingphotondetectionefficiency, lewis_measurements_2025}, which also contains a description of the methodology. $ P_h(\Vov)$ is difficult to measure directly in a p-on-n SiPM, but can be determined using DeCT rates as described below. In cases where this is not possible, the $ P_h(\Vov)$ values derived from photodetection efficiency measurements in \cite{croix2025mappingphotondetectionefficiency} are used.



\subsection{HPK VUV4 SiPM analysis \label{HPK_analysis}}

Delayed cross-talk (DeCT) events in this device, with delay times on the order of nanoseconds, are assumed to originate from photons absorbed in the substrate as drift time for electrons generated in the p-type region to reach the junction region and trigger an avalanche is typically less than 10 ps \cite{Acconcia:23}, and the insensitive region near the surface is considered to be on the order of 1~nm \cite{croix2025mappingphotondetectionefficiency}. This implies that all observable DeCT pulses are triggered by holes originating from the n-type substrate. Consequently, the trend in $P_{h}(V)$ with $\Vov$ can be determined through the $\Vov$-dependent behavior of DeCT events.

\begin{figure}[htbp]
\centering
\begin{subfigure}[b]{0.44\textwidth}
    \centering
    \includegraphics[width=\textwidth]{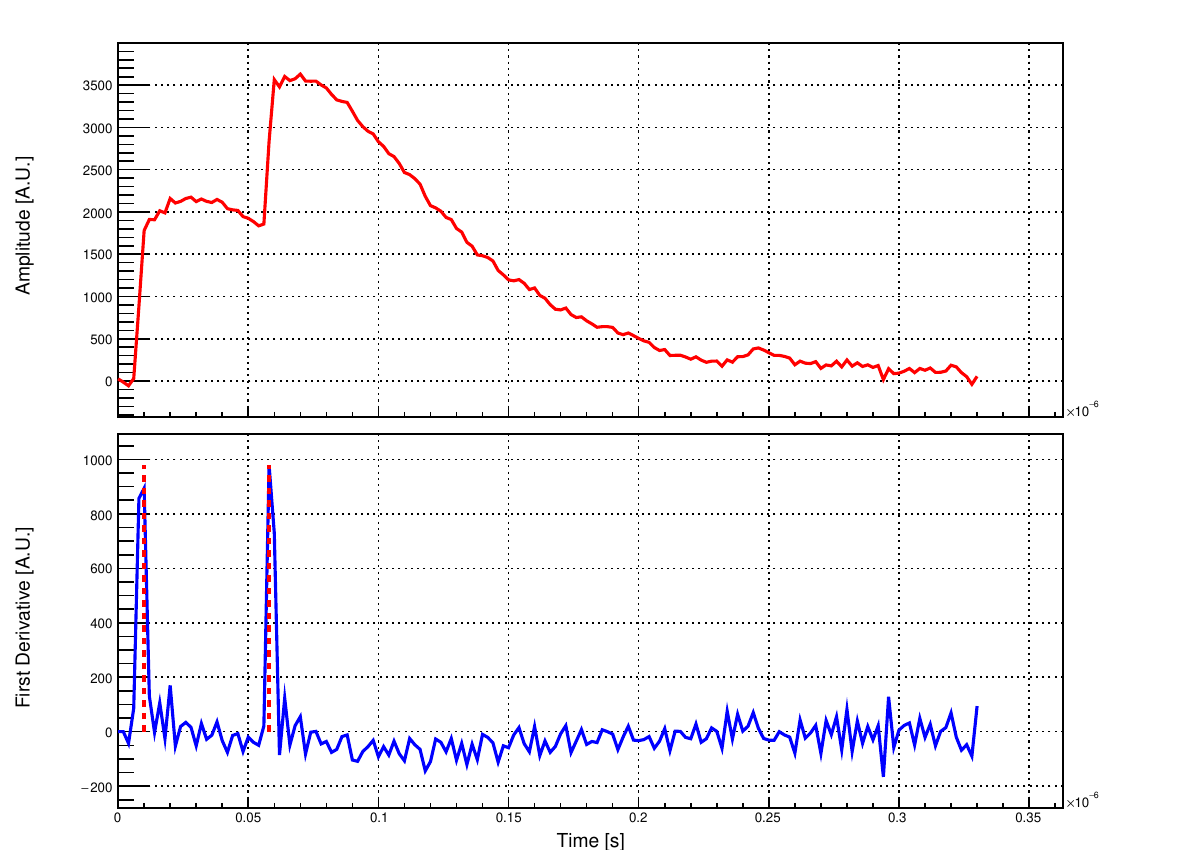}
    \caption{Example of a DeCT raw waveform (top) and its DLED filtered waveform (bottom).}
    \label{fig:sub_DeCT}
\end{subfigure}
\qquad
\begin{subfigure}[b]{0.44\textwidth}
    \centering
    \includegraphics[width=\textwidth]{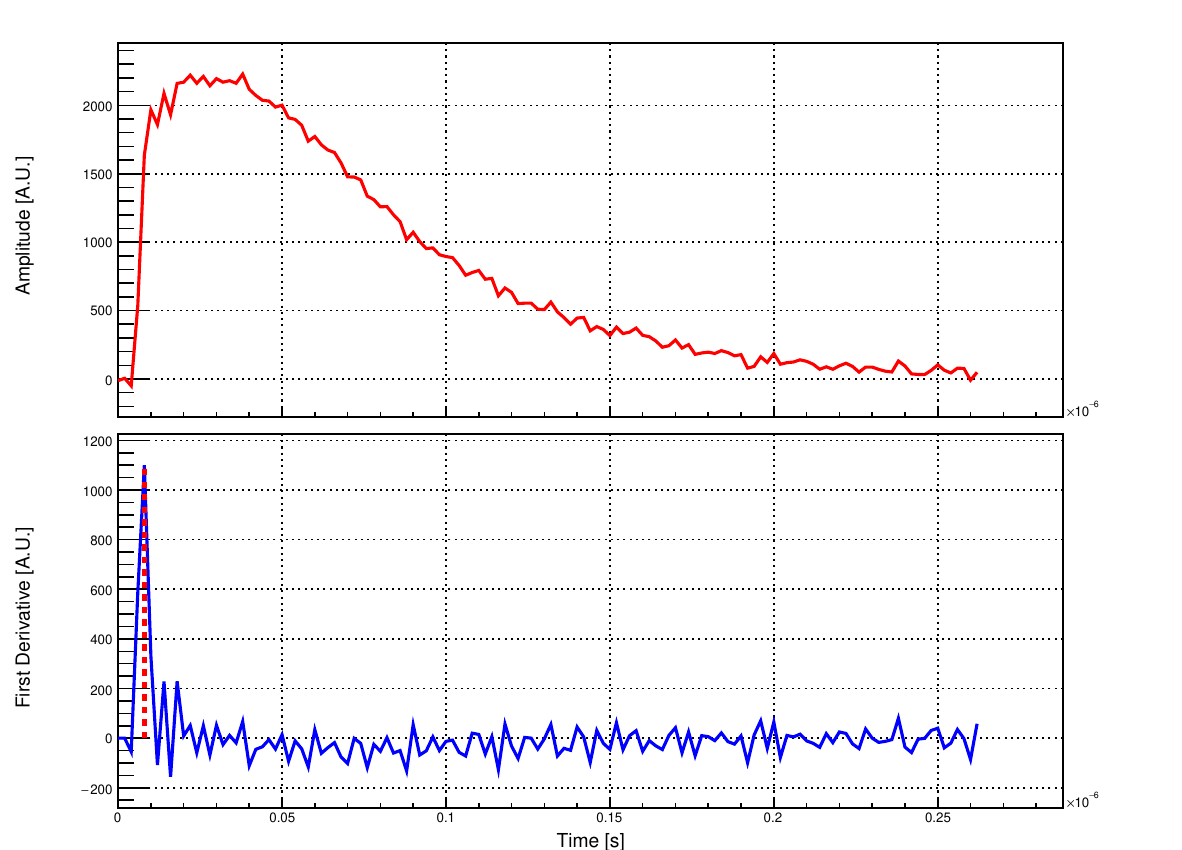}
    \caption{Example of a DiCT raw waveform (top) and its DLED filtered waveform (bottom).}
    \label{fig:sub_DiCT}
\end{subfigure}
\caption{Two examples to show the method of separating DeCT events from DiCT events. The red dotted lines in the DLED filtered waveform mark the timestamps of the peaks.}
\label{fig:CT_example}
\end{figure}

The pulse shapes of DiCT and DeCT events exhibit distinct characteristics that allow for their differentiation. DiCT events will appear indistinguishable from a genuine (prompt) multi-PE pulse. Conversely, the DeCT pulse shape features a time delay of nanoseconds between two or more SPE signals. This distinction enables easy identification of DiCT and DeCT through a DLED (Delayed Leading-Edge Discriminator) filter \cite{article}. As shown in  ~\autoref{fig:CT_example}, DLED filtered waveforms for DeCT consistently produces two prominent, sharp peaks (\autoref{fig:sub_DeCT}), while DiCT yields only a single peak (\autoref{fig:sub_DiCT}). Additionally, by measuring the time interval between the two peaks in DeCT, the crosstalk delay time distribution can be constructed.

\begin{figure}[htbp]
\centering
\includegraphics[width=.52\textwidth]{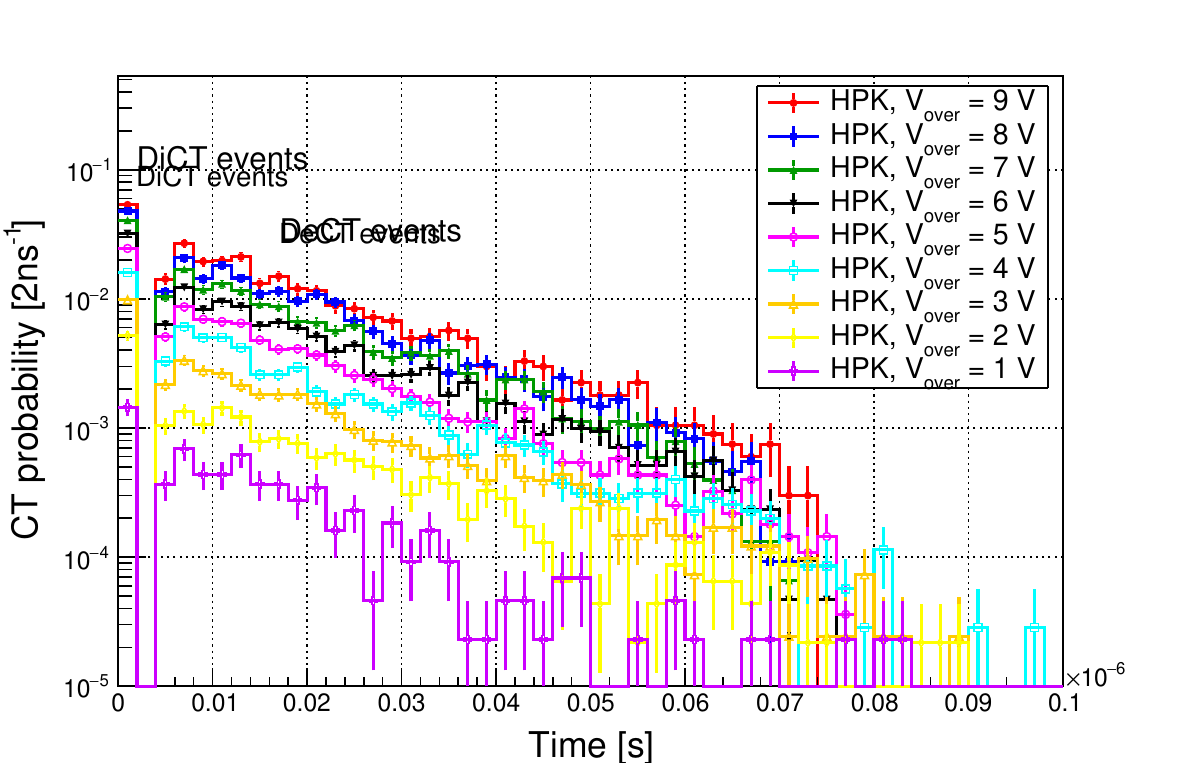}
\caption{The distribution of crosstalk events versus delay time for the Hamamatsu VUV4. As the delay time for DiCT is effectively zero, all DiCT events are shown in the first bin. \label{fig:HPK_DeCT}}
\end{figure}


~\autoref{fig:HPK_DeCT} shows the crosstalk probability in the HPK SiPM as a function of delay time, for $\Vov$s ranging from $\Vov = 1$ to $9~ \text{V}$. To avoid complications from recursive effects—where DiCT and DeCT can coexist in multi-photoelectron signals—only 2PE events were used to construct the probability distributions. The resulting curves in   ~\autoref{fig:HPK_DeCT} can be divided into two distinct regions. The first region corresponds to the initial bin, where the delay time is less than 4~ns and is attributed to DiCT  events. The second region comprises DeCT events with delay times larger than 4 ns, exhibiting a distribution where the probability gradually decreases as the delay time increases.

\begin{figure}[htbp]
\centering
\includegraphics[width=.5\textwidth]{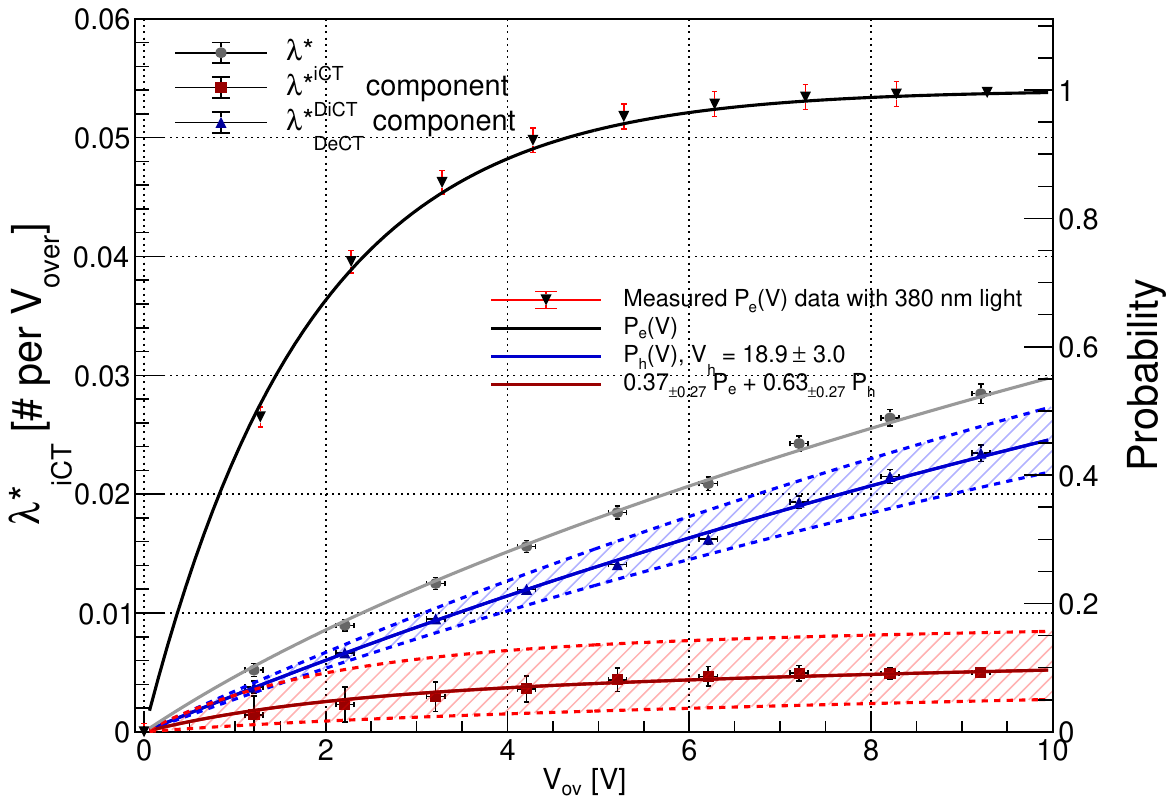}
\caption{$\lambda^*$ values for the Hamamatsu VUV4 at -60°C, shown with $P_e$ and  $P_h$ curves for the same device. The gray data points representing the overall $\lambda^*$, decomposed into $\lambda^*_{DeCT}$ (blue points) and $\lambda^*_{DiCT}$ (red points) The red line shows a fit to $\lambda^*_{DeCT}$ using  ~\autoref{eq:lambdatoV}.}

\label{fig:HPK_CT_component}
\end{figure}


 ~\autoref{fig:HPK_DeCT} shows the expected increase in both  $\lambda^*_{DiCT}$ and $\lambda^*_{DeCT}$ with $\Vov$. The trend of $\lambda^*_{DiCT}$ and $\lambda^*_{DeCT}$ as a function of $\Vov$ is presented in  ~\autoref{fig:HPK_CT_component}. The black data points in  ~\autoref{fig:HPK_CT_component} give the independently measured $P_e(V)$, from which the functional form for $P_e(V)$ is fit. The gray data points represent $\lambda_{iCT}^*$, which is further divided into the delayed and prompt components: $\lambda^*_{DeCT}$ in blue data points and $\lambda^*_{DiCT}$ in red. The functional shape of $P_{h}(V)$ is considered to be equivalent to that of  $\lambda^*_{DeCT}$ as described above. A shape parameter of $V_h = 18.9\pm3.0$~V is fit from this delayed cross-talk data, giving good agreement with the value of $16.62\pm1.2$ determined from fitting PDE data in \cite{croix2025mappingphotondetectionefficiency}. The proportion of DiCT events initiated by each carrier type can then be determined by fitting to $\lambda^*_{DiCT}$ using  ~\autoref{eq:lambdatoV}. Of photons absorbed in the active region, $36 \pm 26\%$ are absorbed in the p-region while the remaining $64 \pm 26\%$ are absorbed in the active n-region. For the HPK VUV4 SiPM, when considering both DiCT and DeCT signals, $7.35 \pm 5.3\%$ of photons may initiate e-driven avalanches, while the remaining $92.6 \pm 5.3\%$ produce holes in either the active n-region or substrate. Because the functions of $P_e(V)$ and $P_h(V)$ for HPK VUV4 SiPM are known, the total probability of electron or hole-driven avalanches from secondary photons can be calculated using either term in \autoref{eq:lambdaiCT}.  \autoref{fig:ratio} illustrates the ratio of electron to hole driven avalanches for all iCT, as a function of $\Vov$.



\subsection{FBK VUV-HD3 SiPM analysis}\label{sec:FBK_Pe}

\begin{figure}[htbp]
\centering
\includegraphics[width=.5\textwidth]{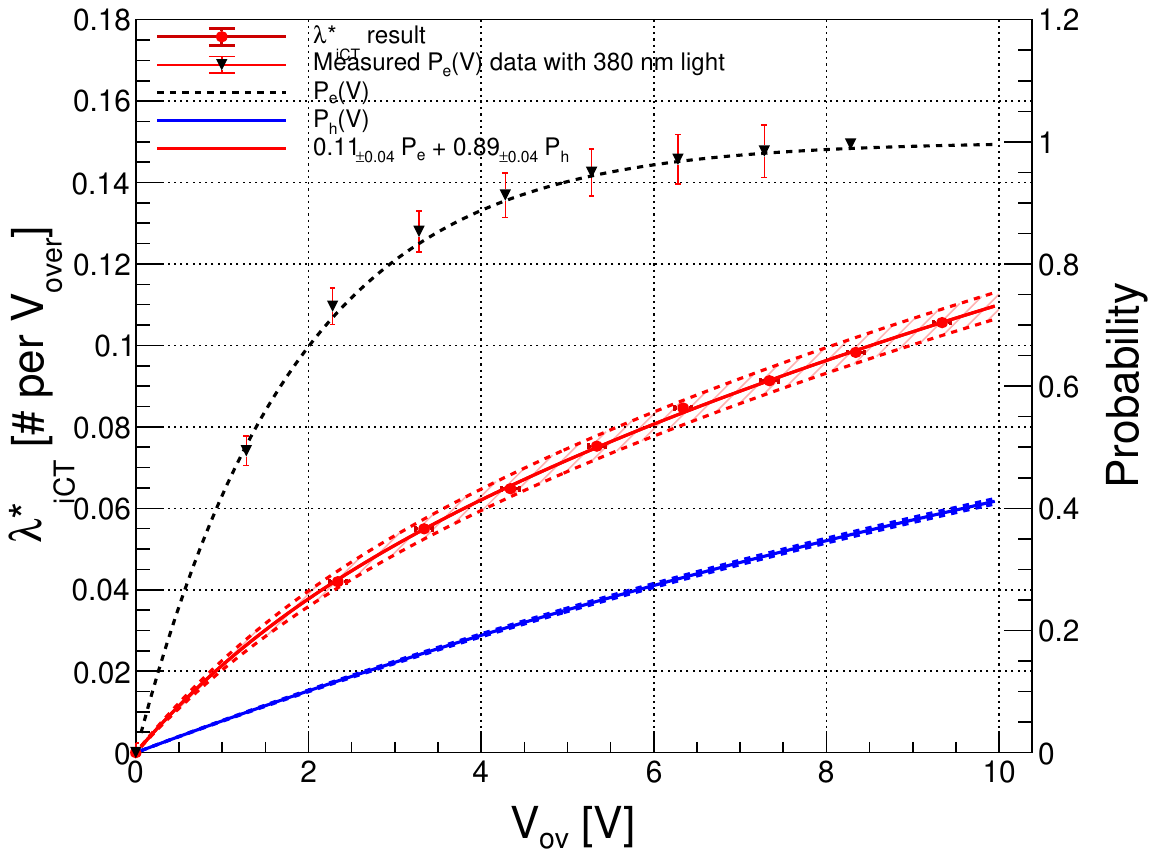}
\caption{$\lambda^*$ values for the FBK VUV-HD3 at -60°C, shown with $P_e$ and  $P_h$ curves for the same device. The red line shows a fit to $\lambda^*$ using  ~\autoref{eq:lambdatoV}. $P_h$ values are from \cite{croix2025mappingphotondetectionefficiency}.}

\label{fig:FBK_Pe}

\end{figure}



Compared to the HPK VUV4 SiPM, the source analysis of the FBK VUV-HD3 SiPM is relatively straightforward as negligible DeCT is observed. The internal crosstalk can be investigated through direct fitting of the $\lambda^*$ data, as shown in  ~\autoref{fig:FBK_Pe}. In the figure, the red data points represent the measured $\lambda^*$ values at –60°C as a function of $\Vov$. A linear combination of $P_e(V)$ and $P_h(V)$, following  ~\autoref{eq:lambdatoV}, is used to fit the data and extract the relative contributions of electron and hole-triggered avalanches to the crosstalk signals. $P_e(V)$ data are from \cite{croix2025mappingphotondetectionefficiency}. As $P_h(V)$ cannot be measured separately due to the absence of DeCT events, the value of $V_h = 18.8 \pm 0.3$ from \cite{croix2025mappingphotondetectionefficiency} is used when fitting for $\alpha$ and $\beta$. 

\begin{figure}[htbp]
\centering
\includegraphics[width=.5\textwidth]{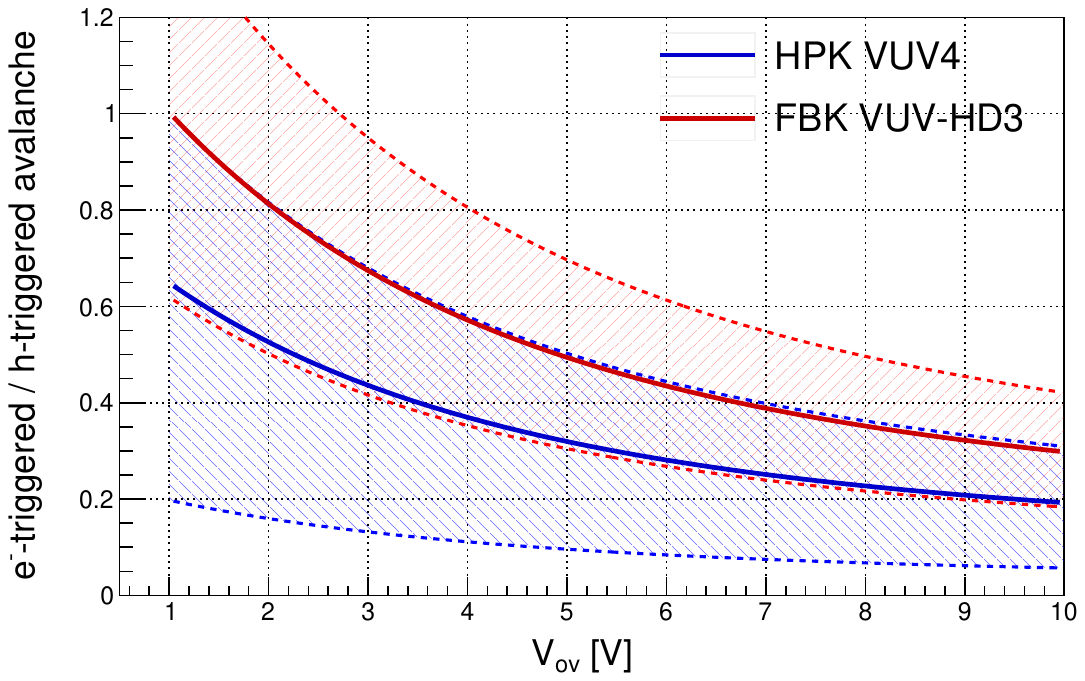}
\caption{ The ratio of number of $e^-$-triggered avalanches to hole-triggered avalanches for all iCT, as a function of $\Vov$ for both devices.}
\label{fig:ratio}
\end{figure}

In  ~\autoref{fig:FBK_Pe}, the black and blue curves represent the $P_e(V)$ and $P_h(V)$ functions, respectively, while the red curve shows the final fit result. 
The fit results indicatet that for the FBK VUV-HD3 SiPM, approximately $11 \pm 4\%$ of iCT photons are absorbed in the p-region with the remaining $89 \pm 4\%$ absorbed in the n-region. Multiplying by $P_e$, $P_h$ gives the number of electron or hole driven events, with the total iCT ratio as function of $\Vov$ given in   \autoref{fig:ratio}. The values of $\alpha$ and $\beta$ for both the HPK VUV4 and FBK VUV-HD3 SiPMs are summarized and compared in ~\autoref{tab1}. As DeCT and DiCT are both significant in the HPK VUV4, and can be separated, $\alpha$ and $\beta$ values for each component are listed individually. Values for the overall crosstalk signal are also given. As discussed above, all DeCT originates from hole diffusion into the active region, meaning that 100\% of DeCT events are hole-initiated avalanches.

\begin{table}
    \centering
    \caption{Photo-absorption fractions $\alpha$ and $\beta$ contributing to different crosstalk sources in FBK VUV-HD3 and HPK VUV4 SiPMs. Uncertainties were obtained from fittings.}\label{tab1}
    \begin{tabular}{l l cc}
        \toprule
        SiPM model & Crosstalk type & $\alpha$ (\%) & $\beta$ (\%)\\
        \midrule
        \multirow{1}{*}{FBK VUV-HD3} & iCT & $11.0 \pm 4.0$ & $89.0 \pm 4.0$ \\
        \midrule
        \multirow{3}{*}{HPK VUV4} & iCT & $7.35 \pm 5.3$ & $92.6 \pm 5.3$ \\
                                    & DeCT & $0$ & $100$ \\
                                 & DiCT & $37.2 \pm 26.8$ & $62.8 \pm 26.8$ \\
                                 
        \bottomrule
    \end{tabular}
\end{table}

\section{Discussion and Conclusions}

In either device, secondary photons are predominantly absorbed in the n doped region or substrate, mainly attributed to a significantly higher volume of material. As can be seen in \autoref{fig:ratio}, other than at very low $\Vov$ where $P_h$ is small, crosstalk is dominated by hole-driven mechanisms as $\beta/\alpha > P_e/P_h$. However, if DiCT alone is considered, the $\alpha$ values observed (11\% and 37.2\% for the FBK and HPK devices respectively) are significantly higher than the modelled proportion of p-doped material in the device volume as calculated in \cite{croix2025mappingphotondetectionefficiency}\footnote{approximated as the ratio of $X_{PN}/dw^*$}. What this indicates is that a higher proportion of DiCT events are electron-triggered than would be observed if photons were absorbed uniformly throughout the device structure, as would be expected if transmission through trenches were the dominant optical path for DiCT.
This suggests that reflection of secondary photons from the device surface to neighboring cells is a significant mechanism for crosstalk, as, although the absorption length at the secondary emission wavelengths is long enough that photons may travel through the full depth of the device, the majority will be absorbed in the initial segment of their path (which will be in the p-region for photons originating from the device surface).

The results of this work illustrate that metallic trenches and reducing hole lifetime via doped substrate are effective for mitigating DiCT and DeCT, respectively. However, it may not be practical to implement these features for all device architectures. Two additional design methodologies are discussed to mitigate crosstalk in future devices:

First, minimize reflections from the device surface by an appropriate choice of anti-reflection coating, so that secondary photons are transmitted out of the SiPM rather than reflecting back into the originating cell or neighbouring cells. However, this may impact detection efficiency or enhance photon emission from the SPAD. Al$_2$O$_3$, fabricated by the deposition of a nanoparticle layer, has been successfully implemented as an anti-reflection coating to improve the sensitivity of photodiodes at the NIR wavelengths relevant to secondary photon emission \cite{ismail_preparation_2017}. An Al$_2$O$_3$ coating applied after etching of native oxide would likely decrease internal crosstalk, as it would decrease the reflectivity of the surface interface at NIR wavelengths. This would result in a higher proportion of secondary photons being transmitted away from the device rather than reflected back into the bulk where they may cause crosstalk avalanches. However, it may increase susceptibility to external crosstalk for SiPMs employed in a larger detector - such a coating could be combined with short-pass filters to mitigate these drawbacks.

Secondly, the dominance of hole-initiated avalanches in crosstalk events implies that crosstalk probability would be reduced by minimizing the ratio of $P_h/P_e$. As $P_h$ and $P_e$ depend respectively on the electron and hole impact ionization coefficients, and the ratio of the electron to the hole impact ionization coefficients is increased at lower electric fields \cite{McIntyre_1973}, it is hypothesized that a device structure operating with a lower peak electric field would demonstrate a lower crosstalk probability. Analogous field engineering techniques have been successfully applied to produce low-noise linear APD structures, as APD noise also depends on the ratio between the electron and hole ionization coefficients \cite{paulus_comparison_1994}. Low-field technology has been demonstrated by FBK to yield improvements in DCR, afterpulsing probability and maximum PDE, but only a modest decrease was observed in DiCT probability compared to standard-field technology \cite{acerbi_cryogenic_2017}. This suggests that further device modeling work is required to study the effects of field engineering on DiCT, which is beyond the scope of this study.

All the experimental data presented here were obtained using p-on-n SiPMs (FBK VUV-HD3 and Hamamatsu VUV4). This choice is not because the proposed method is limited to p-on-n devices, but simply because our current studies focus on these two SiPMs as the candidates of the nEXO experiment. In principle, the same approach can also be applied to n-on-p, VIS-sensitive SiPMs.

We anticipate that this methodology can be employed in identifying and mitigating the dominant sources of crosstalk in novel SiPM structures. This would be particularly effective if a two-photon absorption measurement technique \cite{pape_study_2024} can be applied successfully to SPAD and SiPM devices. This would allow measurement of device response to photons injected anywhere in the structure facilitating the direct measurement of $P_e$ and $P_h$, enabling the presented methodology to be used without the need for measurements of DeCT or external efficiency modeling work. This work also informs understanding of delayed coincidence arising from NIR photon absorption outside the active region, pertaining to external cross-talk between SiPMs. Lastly, the $P_h$ curves measured here validate the results for the `default' fit in \cite{croix2025mappingphotondetectionefficiency}, excluding results yielding higher $P_h$ values and smaller junction size. This resolves degeneracy between model parameters and input photoabsorption data, improving accuracy of future photo-detection efficiency modeling.


\bibliographystyle{ieeetr}
\bibliography{biblio}

\begin{thebibliography}{10}

\bibitem{gallina_performance_2022}
G.~Gallina, Y.~Guan, and F.~Retiere, ``Performance of novel {VUV}-sensitive {Silicon} {Photo}-{Multipliers} for {nEXO},'' {\em European Physical Journal. C, Particles and Fields (Online)}, vol.~82, Dec. 2022.

\bibitem{carnesecchi_light_2020}
F.~Carnesecchi, ``Light detection in {DarkSide}-20k,'' {\em Journal of Instrumentation}, vol.~15, p.~C03038, Mar. 2020.

\bibitem{lecoq_sipm_2021}
P.~Lecoq and S.~Gundacker, ``{SiPM} applications in positron emission tomography: toward ultimate {PET} time-of-flight resolution,'' {\em The European Physical Journal Plus}, vol.~136, p.~292, Mar. 2021.
\newblock Number: 3 Publisher: Springer Berlin Heidelberg.

\bibitem{4907333}
J.~Nissinen, I.~Nissinen, and J.~Kostamovaara, ``Integrated receiver including both receiver channel and tdc for a pulsed time-of-flight laser rangefinder with cm-level accuracy,'' {\em IEEE Journal of Solid-State Circuits}, vol.~44, no.~5, pp.~1486--1497, 2009.

\bibitem{7049394}
J.~Kostamovaara, J.~Huikari, L.~Hallman, I.~Nissinen, J.~Nissinen, H.~Rapakko, E.~Avrutin, and B.~Ryvkin, ``On laser ranging based on high-speed/energy laser diode pulses and single-photon detection techniques,'' {\em IEEE Photonics Journal}, vol.~7, no.~2, pp.~1--15, 2015.

\bibitem{Bruschini2019}
C.~Bruschini, H.~Homulle, I.~M. Antolovic, S.~Burri, and E.~Charbon, ``Single-photon avalanche diode imagers in biophotonics: review and outlook,'' {\em Light: Science \& Applications}, vol.~8, no.~1, p.~87, 2019.

\bibitem{hampel_optical_2020}
M.~R. Hampel, A.~Fuster, C.~Varela, M.~Platino, A.~Almela, A.~Lucero, B.~Wundheiler, and A.~Etchegoyen, ``Optical crosstalk in {SiPMs},'' {\em Nuclear Instruments and Methods in Physics Research Section A: Accelerators, Spectrometers, Detectors and Associated Equipment}, vol.~976, p.~164262, Oct. 2020.

\bibitem{akil_multimechanism_1999}
N.~Akil, S.~Kerns, D.~Kerns, A.~Hoffmann, and J.-P. Charles, ``A multimechanism model for photon generation by silicon junctions in avalanche breakdown,'' {\em IEEE Transactions on Electron Devices}, vol.~46, pp.~1022--1028, May 1999.

\bibitem{mclaughlin_characterisation_2021}
J.~B. McLaughlin, G.~Gallina, F.~Retière, A.~De~St.~Croix, P.~Giampa, M.~Mahtab, P.~Margetak, L.~Martin, N.~Massacret, J.~Monroe, M.~Patel, K.~Raymond, J.~Roiseux, L.~Xie, and G.~Zhang, ``Characterisation of {SiPM} {Photon} {Emission} in the {Dark},'' {\em Sensors}, vol.~21, p.~5947, Jan. 2021.
\newblock Publisher: Multidisciplinary Digital Publishing Institute.

\bibitem{NAGY201444}
F.~Nagy, M.~Mazzillo, L.~Renna, G.~Valvo, D.~Sanfilippo, B.~Carbone, A.~Piana, G.~Fallica, and J.~Molnár, ``Afterpulse and delayed crosstalk analysis on a stmicroelectronics silicon photomultiplier,'' {\em Nuclear Instruments and Methods in Physics Research Section A: Accelerators, Spectrometers, Detectors and Associated Equipment}, vol.~759, pp.~44--49, 2014.

\bibitem{7102791}
F.~Acerbi, A.~Ferri, G.~Zappala, G.~Paternoster, A.~Picciotto, A.~Gola, N.~Zorzi, and C.~Piemonte, ``Nuv silicon photomultipliers with high detection efficiency and reduced delayed correlated-noise,'' {\em IEEE Transactions on Nuclear Science}, vol.~62, no.~3, pp.~1318--1325, 2015.

\bibitem{tajima_studies_2023}
H.~Tajima, A.~Okumura, and K.~Furuta, ``Studies of propagation mechanism of optical crosstalk in silicon photomultipliers,'' {\em Nuclear Instruments and Methods in Physics Research Section A: Accelerators, Spectrometers, Detectors and Associated Equipment}, vol.~1049, p.~168029, Apr. 2023.

\bibitem{croix2025mappingphotondetectionefficiency}
A.~de~St~Croix, H.~Lewis, K.~Raymond, F.~Retière, M.~Henriksson-Ward, G.~Gallina, N.~Morrison, and A.~Zhang, ``Mapping the photon detection efficiency of vuv sensitive sipms from the ultra-violet to the near infra-red,'' 2025, https://arxiv.org/abs/2508.16005.

\bibitem{yamamoto_recent_2019}
K.~Yamamoto, T.~Nagano, R.~Yamada, T.~Ito, and Y.~Ohashi, ``Recent {Development} of {MPPC} at {Hamamatsu} for {Photon} {Counting} {Applications},'' in {\em Proceedings of the 5th {International} {Workshop} on {New} {Photon}-{Detectors} ({PD18})}, vol.~27 of {\em {JPS} {Conference} {Proceedings}}, Journal of the Physical Society of Japan, Nov. 2019.

\bibitem{oleynikov_after-pulsing_2017}
V.~Oleynikov and V.~Porosev, ``After-pulsing and cross-talk comparison for {PM1125NS}-{SB0} ({KETEK}), {S10362}-11-{100C} ({HAMAMATSU}) and {S13360}-{3050CS} ({HAMAMATSU}),'' {\em Journal of Instrumentation}, vol.~12, p.~C06046, June 2017.

\bibitem{merzi_nuv-hd_2023}
S.~Merzi, S.~E. Brunner, A.~Gola, A.~Inglese, A.~Mazzi, G.~Paternoster, M.~Penna, C.~Piemonte, and M.~Ruzzarin, ``{NUV}-{HD} {SiPMs} with metal-filled trenches,'' {\em Journal of Instrumentation}, vol.~18, p.~P05040, May 2023.

\bibitem{ninkovic_avalanche_2007}
J.~Ninković, R.~Eckhart, R.~Hartmann, P.~Holl, C.~Koitsch, G.~Lutz, C.~Merck, R.~Mirzoyan, H.-G. Moser, A.-N. Otte, R.~Richter, G.~Schaller, F.~Schopper, H.~Soltau, M.~Teshima, and G.~Vâlceanu, ``The avalanche drift diode—{A} back illumination drift silicon photomultiplier,'' {\em Nuclear Instruments and Methods in Physics Research Section A: Accelerators, Spectrometers, Detectors and Associated Equipment}, vol.~580, pp.~1013--1015, Oct. 2007.

\bibitem{parellada_monreal_back_2021}
L.~Parellada~Monreal, G.~Paternoster, A.~G. Gola, F.~Acerbi, P.~Bellutti, G.~Borghi, L.~Ferrario, A.~Ficorella, S.~Merzi, M.~Ruzzarin, and A.~Mazzi, ``Back {Side} {Illuminated} {SiPM} from {NIR} to {NUV} light detection,'' p.~ESSDERC 2021, 2021.

\bibitem{van_sieleghem_backside-illuminated_2022}
E.~Van~Sieleghem, G.~Karve, K.~De~Munck, A.~Vinci, C.~Cavaco, A.~Süss, C.~Van~Hoof, and J.~Lee, ``A {Backside}-{Illuminated} {Charge}-{Focusing} {Silicon} {SPAD} {With} {Enhanced} {Near}-{Infrared} {Sensitivity},'' {\em IEEE Transactions on Electron Devices}, vol.~69, pp.~1129--1136, Mar. 2022.

\bibitem{CAPASSO2020164478}
M.~Capasso, F.~Acerbi, G.~Borghi, A.~Ficorella, N.~Furlan, A.~Mazzi, S.~Merzi, V.~Mozharov, V.~Regazzoni, N.~Zorzi, G.~Paternoster, and A.~Gola, ``Fbk vuv-sensitive silicon photomultipliers for cryogenic temperatures,'' {\em Nuclear Instruments and Methods in Physics Research Section A: Accelerators, Spectrometers, Detectors and Associated Equipment}, vol.~982, p.~164478, 2020.

\bibitem{hamamatsu_photonics_vuv-mppc_2017}
{Hamamatsu Photonics}, ``{VUV}-{MPPC} 4th generation ({VUV4}),'' Mar. 2017.

\bibitem{acerbi_cryogenic_2017}
F.~Acerbi, S.~Davini, A.~Ferri, C.~Galbiati, G.~Giovanetti, A.~Gola, G.~Korga, A.~Mandarano, M.~Marcante, G.~Paternoster, C.~Piemonte, A.~Razeto, V.~Regazzoni, D.~Sablone, C.~Savarese, G.~Zappalá, and N.~Zorzi, ``Cryogenic {Characterization} of {FBK} {HD} {Near}-{UV} {Sensitive} {SiPMs},'' {\em IEEE Transactions on Electron Devices}, vol.~64, pp.~521--526, Feb. 2017.

\bibitem{lewis_measurements_2025}
H.~Lewis, M.~Mahtab, F.~Retière, A.~De~St.~Croix, K.~Raymond, M.~Henriksson-Ward, N.~Morrison, A.~Zhang, A.~Capra, and R.~Underwood, ``Measurements of the quantum yield of silicon using {Geiger}-mode avalanching photodetectors,'' {\em The European Physical Journal C}, vol.~85, p.~214, Feb. 2025.

\bibitem{Raymond_stimulated_emission_TED_2024}
K.~Raymond, F.~Retière, H.~Lewis, A.~Capra, D.~McCarthy, A.~d.~S. Croix, G.~Gallina, J.~McLaughlin, J.~Martin, N.~Massacret, P.~Agnes, R.~Underwood, S.~Koulosousas, and P.~Margetak, ``Stimulated secondary emission of single-photon avalanche diodes,'' {\em IEEE Transactions on Electron Devices}, vol.~71, no.~11, pp.~6871--6879, 2024.

\bibitem{gallina_characterization_2019_HPK}
G.~Gallina, P.~Giampa, and F.~Retière, ``Characterization of the {Hamamatsu} {VUV4} {MPPCs} for {nEXO},'' {\em Nuclear Instruments and Methods in Physics Research Section A: Accelerators, Spectrometers, Detectors and Associated Equipment}, vol.~940, pp.~371--379, Oct. 2019.

\bibitem{gola_nuv-sensitive_2019}
A.~Gola, F.~Acerbi, M.~Capasso, M.~Marcante, A.~Mazzi, G.~Paternoster, C.~Piemonte, V.~Regazzoni, and N.~Zorzi, ``{NUV}-{Sensitive} {Silicon} {Photomultiplier} {Technologies} {Developed} at {Fondazione} {Bruno} {Kessler},'' {\em Sensors}, vol.~19, p.~308, Jan. 2019.
\newblock Publisher: Multidisciplinary Digital Publishing Institute.

\bibitem{Rech:08}
I.~Rech, A.~Ingargiola, R.~Spinelli, I.~Labanca, S.~Marangoni, M.~Ghioni, and S.~Cova, ``Optical crosstalk in single photon avalanche diode arrays: a new complete model,'' {\em Opt. Express}, vol.~16, pp.~8381--8394, Jun 2008.

\bibitem{7322059}
F.~Acerbi, A.~Ferri, G.~Zappala, G.~Paternoster, A.~Gola, N.~Zorzi, and C.~Piemonte, ``Technological and design improvements of fbk nuv silicon-photomultipliers,'' in {\em 2015 Fotonica AEIT Italian Conference on Photonics Technologies}, pp.~1--3, 2015.

\bibitem{lolx_Gallacher2025}
D.~Gallacher, A.~de~St.~Croix, {\em et~al.}, ``Characterization of external cross-talk from silicon photomultipliers in a liquid xenon detector,'' {\em The European Physical Journal C}, vol.~85, no.~6, p.~692, 2025.

\bibitem{gallina_characterization_2019}
G.~Gallina and F.~. Retière, ``Characterization of {SiPM} {Avalanche} {Triggering} {Probabilities},'' {\em IEEE Transactions on Electron Devices}, vol.~66, pp.~4228--4234, Oct. 2019.
\newblock Conference Name: IEEE Transactions on Electron Devices.

\bibitem{Ghioni1996}
M.~Ghioni, S.~Cova, F.~Zappa, and C.~Samori, ``Compact active quenching circuit for fast photon counting with avalanche photodiodes,'' {\em Review of Scientific Instruments}, vol.~67, pp.~3440 -- 3448, 11 1996.

\bibitem{m_biroth_p_achenbach_w_lauth_a_thomas_analytical_2018}
A.~Thomas, W.~Lauth, P.~Achenbach, and M.~Biroth, ``An analytical approach to predict fundamental cryogenic properties of silicon photomultipliers.,'' in {\em ICASIPM}, 2018.

\bibitem{Acconcia:23}
G.~Acconcia, F.~Ceccarelli, A.~Gulinatti, and I.~Rech, ``Timing measurements with silicon single photon avalanche diodes: principles and perspectives,'' {\em Opt. Express}, vol.~31, pp.~33963--33999, Oct 2023.

\bibitem{article}
A.~Gola, C.~Piemonte, and A.~Tarolli, ``The dled algorithm for timing measurements on large area sipms coupled to scintillators,'' {\em IEEE Transactions on Nuclear Science - IEEE TRANS NUCL SCI}, vol.~59, pp.~358--365, 04 2012.

\bibitem{ismail_preparation_2017}
R.~A. Ismail, S.~A. Zaidan, and R.~M. Kadhim, ``Preparation and characterization of aluminum oxide nanoparticles by laser ablation in liquid as passivating and anti-reflection coating for silicon photodiodes,'' {\em Applied Nanoscience}, vol.~7, pp.~477--487, Oct. 2017.

\bibitem{McIntyre_1973}
R.~McIntyre, ``On the avalanche initiation probability of avalanche diodes above the breakdown voltage,'' {\em IEEE Transactions on Electron Devices}, vol.~20, no.~7, pp.~637--641, 1973.

\bibitem{paulus_comparison_1994}
M.~Paulus, J.~Rochelle, and D.~Binkley, ``Comparison of the beveled-edge and reach-through {APD} structures for {PET} applications,'' in {\em Proceedings of 1994 {IEEE} {Nuclear} {Science} {Symposium} - {NSS}'94}, vol.~4, pp.~1864--1868 vol.4, Oct. 1994.

\bibitem{pape_study_2024}
S.~Pape, M.~Fernández~García, M.~Moll, and M.~Wiehe, ``Study of {Neutron}-, {Proton}-, and {Gamma}-{Irradiated} {Silicon} {Detectors} {Using} the {Two}-{Photon} {Absorption}–{Transient} {Current} {Technique},'' {\em Sensors}, vol.~24, p.~5443, Jan. 2024.
\newblock Number: 16 Publisher: Multidisciplinary Digital Publishing Institute.

\end{thebibliography}

\end{document}